\documentclass[man]{apa6}

\usepackage[american]{babel}
\usepackage{csquotes}
\usepackage[style=apa,sortcites=true,sorting=nyt]{biblatex}
\usepackage{graphicx,epsfig}
\usepackage{amsmath}
\usepackage{amssymb}
\usepackage{color}
\usepackage[makeroom]{cancel}
\usepackage{verbatim}
\usepackage{bm}
\usepackage{fancybox}
\usepackage{color}
\usepackage{mdframed}
\usepackage{tikz}
\usepackage{caption}
\usepackage{subcaption}
\usepackage{threeparttable}
\usepackage{listings} 
\usetikzlibrary{shapes.geometric, arrows} 
\usepackage{multirow}
\usepackage{threeparttable}
\usepackage{hyperref}
\hypersetup{
    colorlinks=true,    
    linkcolor=blue,    
    citecolor=blue,    
    filecolor=blue,    
    urlcolor=blue      
}

\DeclareLanguageMapping{american}{american-apa}
\title{Are the Signs of Factor Loadings Arbitrary in Confirmatory Factor Analysis? Problems and Solutions}
\shorttitle{Signs of Factor Loadings}
\author{Dandan Tang, Steven M. Boker, Xin Tong
}
\affiliation{Department of Psychology, University of Virginia
}
\authornote{This research was supported by the National Institutes of Health under Award Number 4R37HD058305-27. Correspondence concerning this article should be addressed to Dandan Tang, Department of Psychology, University of Virginia. Email: gux8df@virginia.edu
}
\abstract{
The replication crisis in social and behavioral sciences has raised concerns about the reliability and validity of empirical studies. While research in the literature has explored contributing factors to this crisis, the issues related to analytical tools have received less attention. This study focuses on a widely used analytical tool - confirmatory factor analysis (CFA) - and investigates one issue that is typically overlooked in practice: accurately estimating factor-loading signs. Incorrect loading signs can distort the relationship between observed variables and latent factors, leading to unreliable or invalid results in subsequent analyses. Our study aims to investigate and address the estimation problem of factor-loading signs in CFA models. Based on an empirical demonstration and Monte Carlo simulation studies, we found current methods have drawbacks in estimating loading signs. To address this problem, three solutions are proposed and proven to work effectively. The applications of these solutions are discussed and elaborated.
} 
\begin{document} 
\maketitle

\section{Introduction}
The replication crisis in psychology has aroused great attention to the lack of reproducibility in empirical studies across social and behavioral sciences, as well as in other scientific disciplines. This raises concerns about the reliability and validity of empirical research within these fields (Cockburn et al., 2020; Oberauer \& Lewandowsky, 2019). Traditionally, replication, the practice of repeating a study to confirm its finding, is the bedrock of scientific validation. Recently, however,  more and more previously accepted findings have proven difficult or impossible to replicate (Camerer et al., 2018; Open Science Collaboration, 2015;  Youyou et al., 2023). Open Science Collaboration (2015)  found a replication rate of 36\% among 97 experiments from papers published in 2008 in three high-ranking psychology journals. Nosek et al. (2022) found only 64\% of  307 experiments replicated. Even some famous studies, like the social psychological study "elderly-walking" conducted by social psychologist John Bargh and colleagues, and its relevant studies failed to replicate, casting doubts on the theory of the goal priming effect (Bargh, Chen,  \& Burrows, 1996; Harris et al., 2013; Muthukrishna \& Henrich, 2019; Pashler et al., 2012). 

Many studies have investigated the problem of replication, identifying factors that may contribute to it, including p-hacking or Cherry-Picking results, publication bias, low statistical power, questionable research practices, inadequate research training, failure to share data and methods, pressure to publish, complexity of scientific research, etc. For instance, $p$-hacking, manipulating data of statistical analyses to achieve a significant result ($p < .05$; Crane, 2018), may lead to non-replicable research outcomes. A researcher may analyze data in various ways but only report a significant result or stop collecting data as soon as a significant result is found. Klein et al. (2018) discovered that among 28 classic and contemporary published findings, 54\% of the replications had statistically significant results with a significance level of p < .05. Publication bias means that journals tend to publish papers with significant/positive results rather than non-significant/negative results (Wagner III, 2022). This leads researchers to pursue significant outcomes,  sometimes regardless of rigorous methodology. Franco, Malhotra, and Simonovits (2014) found only 20\% of the studies published in social sciences journals reported non-significant results and 60\% of the studies discovered non-significant results but never reported them. In practice, although a study may yield significant findings, the associated statistical power could be low (Anderson \& Maxwell, 2017), meaning there is a high probability of failing to reject a false null hypothesis in the future. A meta-analysis of 44 reviews of statistical power observed a mean statistical power of 0.24 to detect a small effect size (d = .20) with a type I error rate of $\alpha = .05$ (Smaldino \& McElreath 2016).

Although many factors leading to the replication crisis have been investigated, few studies have considered the issues with analytical tools (Van Lissa et al., 2021). Analytical tools, comprising statistical methods and software, are the backbone of empirical research, allowing scientists to discover patterns, make inferences, and establish relationships among variables based on the collected data (Ali \& Bhaskar, 2016; Wen et al., 2018). Reliable and consistent use of analytical tools ensures robust research outcomes. If a study uses flawed or inappropriate statistical methods, its findings could be artifacts of the analysis rather than true reflections of underlying phenomena (Tang \& Wen 2020; Wen et al., 2019).

In this study, we focus on one of the most popular analytical tools in social and behavioral sciences, confirmatory factor analysis (CFA), which has been implemented in every structural equation modeling (SEM) program,  such as {Mplus}, {lavaan}, and {OpenMx} (Asparouhov \& Muthen, 2007; Boker et al., 2021; Rosseel, 2012). This technique is particularly important in the development and validation of measurement instruments, such as psychological tests and surveys because it can test theoretical expectations about the relationships between variables and validate the structure of a test or survey (Lin et al., 2020). Based on CFA models, we can test measurement invariance to check whether the measurement of the psychological construct is invariant across groups or times (Vandenberg \& Lance, 2000). CFA is broadly applied in practice, as reflected in the growing number of studies in databases. A search for "confirmatory factor analysis" in the PsychINFO database yields 456 articles from 1961-1990, 8820 articles from 1991-2010, and 23,652 articles from 2011 to now. 

Despite CFA's popularity and well-established methodological foundations, there are still potential problems with this analytical tool. When testing the measurement invariance using CFA, one important step is to check factor loading invariance, which requires that factor loadings are the same across groups or time. For instance, it is discussed in the literature that the Children's Depression Inventory (CDI) is valid to have a one-factor structure with CFA (Stumper et al., 2019) among adolescents, and the factor loadings of the one-factor structure are invariant between 13-year-old and 16-year-old adolescents. However, when we replicated the study to test whether the structure of CDI is invariant in the same data but without missingness, the factor loadings in the 13-year-old group were all negative, but in the 16-year-old group, they were all positive. The absolute values of the corresponding loadings between the two groups were very close. However, according to the definition, we were not able to conclude that these loadings are invariant between the two groups. The natural question is, do factor loading signs matter or not? This question serves as the motivation for our study.

In CFA, factor loadings represent the strength and direction of the relationship between observed variables and latent factors. A positive loading indicates that the observed score increases as the latent factor score increases. A negative loading indicates that the observed variable score decreases as the latent factor score increases. Thus, an incorrect sign can misinterpret the relationship between an observed variable and the latent factor. If the loading sign in a CFA model is inconsistent or "incorrect" (e.g., not in line with theoretical expectations or prior findings), subsequent analyses based on this model can produce unreliable or invalid results. Furthermore, suppose other researchers attempt to replicate the study and find different signs of the factor loadings or can't reproduce the results, they may not trust the original findings. 

This article aims to explore and address the problem of factor loading signs in CFA. The outline of this article is as follows. We begin with a real-data example to demonstrate the factor loading sign problem in CFA models. Next, we will briefly introduce the estimation method for CFA models, investigate the reasons for the sign reversal problem, and propose solutions. Thereafter, two simulations will be conducted to evaluate the impact of the problem and the effectiveness of these solutions. Then, the empirical example will be re-analyzed to illustrate these solutions in practice. Last, the article discusses the prevalence of the factor loading sign problem and provides practical recommendations for researchers.

\section{An Empirical Example}
In this section, we provide a detailed illustration of the estimation problem in a CFA model using a real data example from the study conducted by Stumper et al. (2019). The factor structure of the CDI was investigated using a sample of 227 adolescents aged approximately 13 at baseline (T1) and 16 at follow-up (T2). The CDI has 27 items, scaling 0-2. Stumper et al. (2019) used a weighted least square mean and variance (WLSMV) estimation method in \textit{Mplus} (Muthén \& Muthén, 2017) to analyze the data because the observed variables were binary. To replicate the analysis, we first analyzed the full data, including missing values, using diagonally weighted least squares (DWLS; Rosseel, 2012) in the \textit{lavaan} package in R (see the codes in our \href{https://github.com/DandanTang0/sign-of-factor-loading/tree/main/example}{GitHub}
), which in \textit{Mplus} is given by the WLSMV estimators (Asparouhov \& Muthen, 2007). As shown in Table 1, the loadings from the one-factor model at both T1 and T2 were positive. However, after removing 12 subjects who did not finish the CDI at age 16,  the loadings at T1 became negative, but at T2, they were still positive.  

When testing measurement invariance, we could not conclude that these loadings are invariant between T1 and T2 because, for the data containing missing values, the loadings are in the same direction, whereas for the data without missingness, they have different signs. Furthermore, the data with missingness and without missingness at T1 are from the same population, but the signs of the loadings are opposite. This indicates that results from a later analysis may not confirm the previous analysis. Researchers may even suspect that removing the 12 subjects with missing values changed the sample distribution.

\begin{table}[ht]
	\centering
	\caption{Factor loadings for a one-factor model of the CDI at Time 1 and Time 2}
	\label{t1} 
	\begin{threeparttable}
	\begin{tabular}{rrrrr}
		\hline
        \multirow{2}*{Item}  & \multicolumn{2}{c}{With missingness}  & \multicolumn{2}{c}{Without missingness} \\ 
        \cline{2-5}
             & T1 & T2 & T1 & T2\\ 
             \hline
		CDI1 &  0.772 & 0.746 & -0.773 & 0.742 \\ 
		CDI2 &   0.649 &  0.649 & -0.639 & 0.660\\ 
		CDI3 &  0.618 &  0.659 & -0.614 & 0.655\\ 
            CDI4 &  0.534 &  0.716 & -0.525 &  0.741\\ 
            CDI5 &  0.552 & 0.692 & -0.567 & 0.688 \\ 
            CDI6 &  0.477 & 0.544 & -0.464 & 0.538\\ 
            CDI7 &  0.747 & 0.829 & -0.747 & 0.826\\ 
            CDI8 &  0.604 & 0.740 & -0.637 & 0.761\\ 
            CDI9 &  0.675 & 0.741 & -0.690 & 0.738\\ 
            CDI10 &  0.822 &  0.770 & -0.823 & 0.766\\ 
            CDI11 &  0.769 & 0.708 & -0.765 & 0.703\\ 
            CDI12 &  0.571 &  0.551 & -0.577 &  0.546\\ 
            CDI13 &  0.503 &  0.626 & -0.532 & 0.618\\ 
            CDI14 &  0.536 &  0.700 & -0.541 &  0.696 \\ 
            CDI15 &  0.451 &  0.576 & -0.457 & 0.580\\ 
            CDI16 &  0.650 &  0.638 & -0.644 & 0.631\\ 
            CDI17 &  0.692 &  0.578 & -0.694 & 0.569\\ 
            CDI18 &  0.416 &  0.554 & -0.404 & 0.559\\ 
            CDI19 &  0.400 &  0.462 & -0.399 & 0.455\\ 
            CDI20 &  0.819 & 0.827 & -0.814 & 0.824 \\ 
            \vdots & \vdots & \vdots & \vdots & \vdots\\ 
            CDI27 &  0.673 & 0.714 & -0.691 & 0.711\\ 
		\hline 
	\end{tabular}
	\begin{tablenotes}
        \footnotesize{Note: The table only lists part of the results because of limited space. For complete results, please see our \href{https://github.com/DandanTang0/sign-of-factor-loading/tree/main/table}{GitHub}.
        }
      \end{tablenotes}
     \end{threeparttable}
\end{table}

\section{Estimation Method}
We will first introduce the estimation method for CFA models to understand the problem of the factor loading signs. If the observed variables $\mathbf{x}$ are continuous, a CFA model can be expressed as: 
\begin{equation}
	\mathbf{x} = \mathbf{\tau} + \Lambda \mathbf{\eta} + \mathbf{\epsilon},\notag
\end{equation}
where $\Lambda$ is a vector of factor loadings, $\mathbf{\eta}$ is a vector of latent factors, $\mathbf{\tau}$ is a vector of  latent intercepts, and $\mathbf{\epsilon}$ is a vector of measurement errors. It is assumed that the measurement error is normally distributed with mean $\mathbf{0}$ and variance $\mathbf{\Theta}$, namely, $\mathbf{\epsilon} \sim \mathcal{N}(\mathbf{0}, \mathbf{\Theta})$.  

If the observed variables $\mathbf{x}$ are categorical and have $C$ categories,   the CFA model can be expressed as: 
\begin{equation}
	\mathbf{x^{\ast }} = \mathbf{\tau} + \Lambda \mathbf{\eta} + \mathbf{\epsilon},\notag
\end{equation}
where $\mathbf{x^{\ast }}$ is an underlying continuous variable that is related to $\mathbf{x}$ though a set of $C + 1$ thresholds, $\mathbf{v} = (v_0, v_1, ..., v_{C+1})$, and $v_0 = - \infty$ and $v_{C+1} = \infty$. The probability of $\mathbf{x} = c$ is given as
\begin{equation}
	p(\mathbf{x} = c)=p( v_c \leq  x^\ast \leq  v_{c+1}),\notag
\end{equation}
where $c= 0, 1, ..., C$.  The covariance structure of the CFA model is 
\begin{equation}
	 \mathbf{\Sigma= \Lambda \Phi \Lambda' + \Theta},
\end{equation}
where $\mathbf{\Sigma}$ is the covariance matrix implied by the CFA model, and $\Phi$ is the covariance of the latent factors (Liu et al., 2022a). 

Maximum Likelihood Estimation (MLE) is often applied to estimate model parameters when the observed variables are continuous and normally distributed. MLE aims to find the parameter values that maximize the likelihood function, meaning the parameter values maximize the probability of observing the current sample data (Li, 2016; Tang \& Tong, 2023). In the CFA model, the likelihood function is given as
\begin{equation}
	\mathbf{F} = ln|\mathbf{\Sigma}| - ln|\mathbf{S}| + tr(\mathbf{S}\mathbf{\Sigma}^{-1}) - p\notag, 
\end{equation}
where $p$ is the number of the observed variables, and $\mathbf{S}$ is the covariance matrix of the observed variables if the observed variables $\mathbf{x}$ are continuous (Liu et al., 2022b), or is the covariance matrix of  $\mathbf{x^{\ast }}$ if the observed variables $\mathbf{x}$ are categorical.

When the observed variables are categorical, weighted least square (WLS) estimation is often applied to estimate model parameters. WLS aims to find the parameter values that minimize the fit function, meaning the parameter values minimize the difference between the observed data and the theoretical model based on a weight matrix (Asparouhov \& Muthen, 2007; Li, 2016). In the CFA model, the fit function can be expressed by
\begin{equation}
	\mathbf{F_{wls}} = (s-\sigma(\theta)')\mathbf{W^{-1}}(s-\sigma(\theta) ),\notag
\end{equation}
where $\theta$ is the vector of model parameters, $\mathbf{W}$ is the weight matrix, $\sigma(\theta)$ is the model-implied vector containing the nonredundant, vectorized elements of $\mathbf{\Sigma}$, and $s$ is the vector containing the unique elements of sample statistics (i.e., threshold and polychoric correlation estimates; Li, 2016). If $\mathbf{W}$ is a diagonal matrix, where off-diagonal entries are 0,  and diagonal entries remain the same, WLS will become WLSMV (Asparouhov \& Muthen, 2007). 
                      
\subsection{Problems and Solutions}
Because the latent factor itself does not have a natural scale, Equation (1) is not identified. To identify it, two approaches are often used in popular software for CFA, such as \textit{lavaan}, \textit{Mplus}, and \textit{OpenMx} (Asparouhov \& Muthen, 2007; Boker et al., 2021; Rosseel, 2012). One approach is to fix the factor variance $\sigma^2_{\eta}$, a diagonal element of the covariance matrix $\Phi$, to 1 and the factor means $\mu_{\eta}$ to 0, called fixed factor variance. Another approach is to fix one of the factor loadings on each factor to 1, called fixed factor loading.   

From Equation (1),  we obtain   \begin{equation}
	 cov(x_{i},x_{j}) ~\mbox{or}~ cov(x_{i}^{\ast },x_{j}^{\ast }) = \lambda_i\lambda_j\sigma^2_{\eta} + \sigma^2_{ij},
\end{equation}
where $i = 1, ..., p$ and $j = 1, ..., p$.  Without loss of generality, we assume that when $i=j$,  $\sigma^2_{ij} = \sigma^2_{i}$; otherwise,  $\sigma^2_{ij} = 0$. When using the first fixing method to identify the model, Equation (2) can be written as 
 \begin{equation}
	 cov(x_{i},x_{j}) ~\mbox{or}~ cov(x_{i}^{\ast },x_{j}^{\ast }) = \lambda_i\lambda_j + \sigma^2_{ij}.
\end{equation}
If the covariance $cov(x_{i},x_{j})~\mbox{or}~ cov(x_{i}^{\ast },x_{j}^{\ast })$ is positive, Equation (3) indicates that $\lambda_i$ and $\lambda_j$ should be both positive or negative. Whether positive or negative, the loadings are correctly estimated,  but the direction of the estimated loadings could be opposite. This explains why the loadings at T1 from the data without missing values are all negative in the empirical example. However, when factor loading signs are opposite, the interpretation of the factor also has opposite meanings. 

When using the second fixing method to identify the model,  Equation (2) can be written as 
 \begin{equation}
	 cov(x_{1},x_{j}) ~\mbox{or}~ cov(x_{1}^{\ast },x_{j}^{\ast }) = \lambda_j\sigma^2_{\eta} + \sigma^2_{1j}.
\end{equation}
Equation (4) indicates that  $\lambda_j$ should keep the same sign as the covariance $cov(x_{1},x_{j})~\mbox{or}~ cov(x_{1}^{\ast },x_{j}^{\ast })$. However, if the true value of the first loading is negative but fixed to 1,  this will also lead to a direction problem in the estimation. To solve this estimation problem, we propose to fix a positive loading to 1, which can help other loadings to be of the expected sign. 

To fit a CFA model using  MLE or WLS, an optimization algorithm iteratively adjusts the parameter estimates to find the values that maximize the likelihood or minimize the fit function (Kochenderfer \& Wheeler, 2019). As shown by the symmetric curve in Figure 1,  the optimal solution can fall in the negative or positive range of the x-axis. To solve this estimation problem, we propose to set the lower or upper bounds of the loadings to be larger (smaller) than 0 when the true values of the loadings are expected to be larger (smaller) than 0. 

\begin{figure}[h]
    \centering
    \includegraphics[scale=0.9]{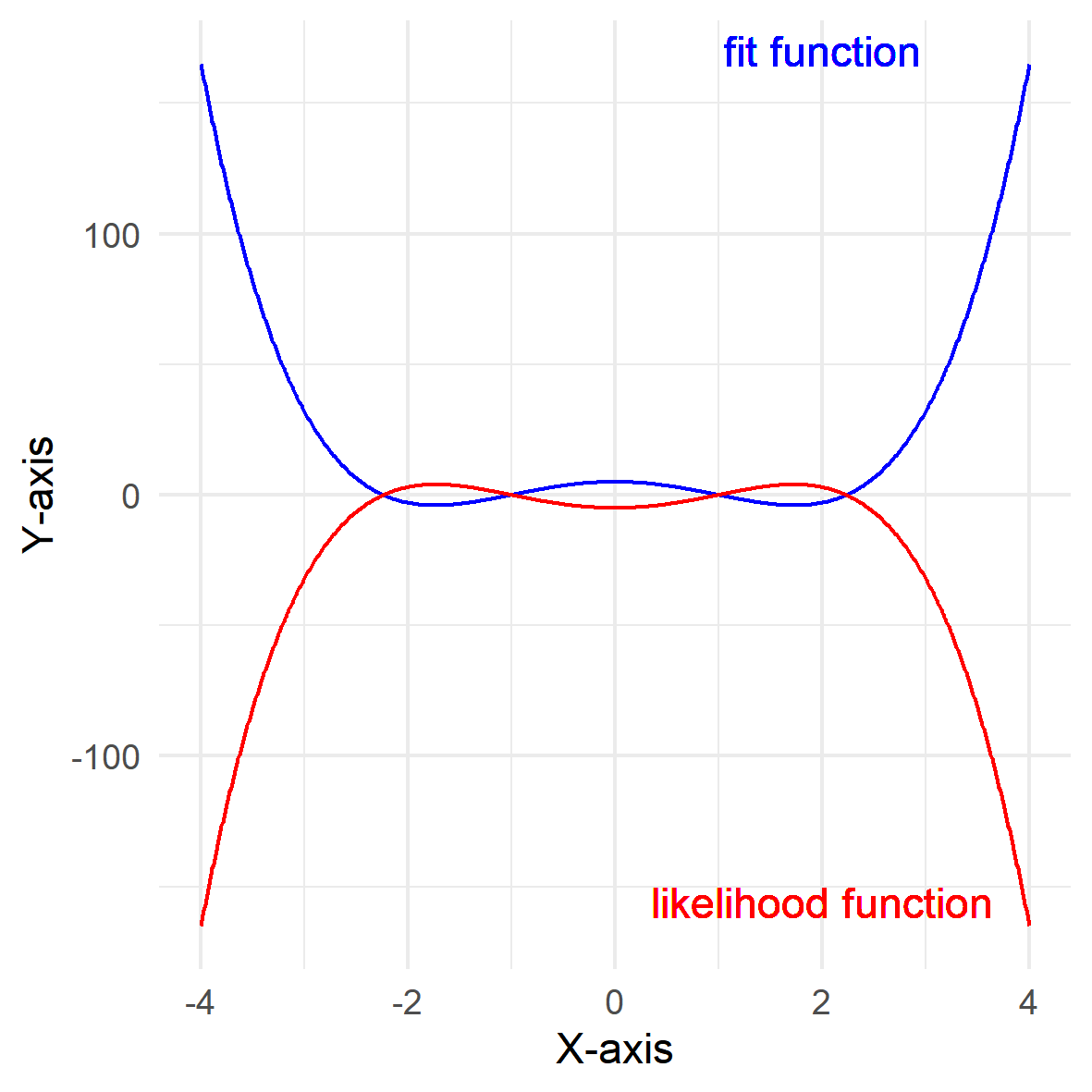} 
    \caption{ The sample graph of the likelihood and fit function 1 } 
\end{figure}

Furthermore, the optimization algorithm needs a starting value for each parameter. The starting value is an initial guess for the parameter, and then the estimation procedure iteratively refines this starting guess to converge toward the parameter's most optimal value (Kochenderfer \& Wheeler, 2019). Sometimes, algorithms can get stuck in "local" solutions, which are specific to the region around the starting values, rather than finding the "global" solution, which is the best solution overall (Arora et al., 1995). Figure 2 shows parameter estimates can be negative when algorithms get stuck in "local" solutions at the negative x-axis. To solve this estimation problem,  we propose adjusting starting values to obtain factor loading estimates with expected signs. In addition, a large positive or small negative starting value may help obtain positive or negative estimates. 

\begin{figure}[h]
    \centering
    \includegraphics[scale=0.9]{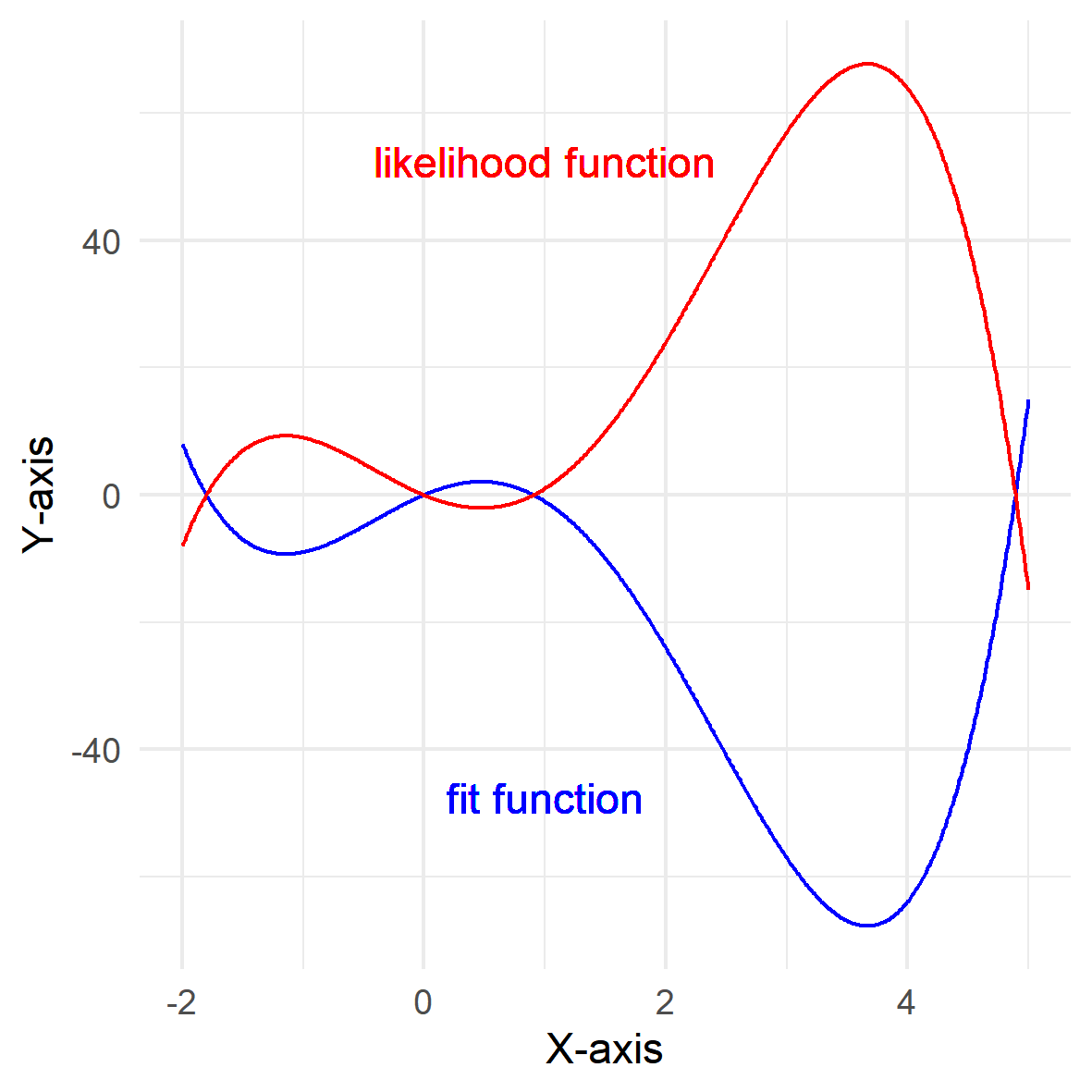} 
    \caption{ The sample graph of the likelihood and fit function 2} 
\end{figure}

\begin{figure}
	\centering
		\includegraphics[scale=0.8]{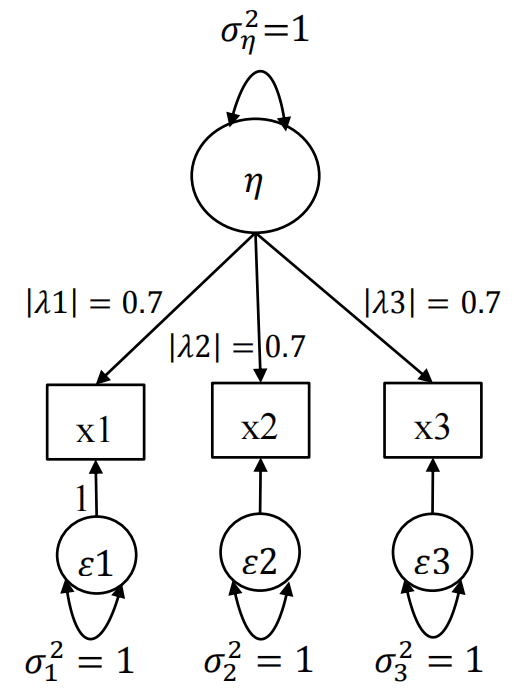}
	
	\caption{Path diagram of a confirmatory factor analysis model.}                 
\end{figure}
We conducted two Monte Carlo simulations to understand the estimation problem and proposed solutions comprehensively. The first simulation study will assess how different estimation methods and computational algorithms affect the signs and consistency of factor loadings across the above CFA software under various situations. The second simulation study will further assess our solutions to the estimation problem.
\section{Simulation 1: Investigating the Estimation Problem}
\subsection{Simulation Design}
We conducted a Monte Carlo simulation study to systematically investigate the estimation problems with two different methods in CFA for model identification. Regardless of continuous or categorical observed variables, the CFA model has the same estimation problem. Thus, the simulation focuses on continuous observed variables using MLE. 

A one-factor CFA model with three continuous indicators was specified as the population model (see Figure 3) from which the data were generated. In the population model, the absolute values of all the loadings $|\lambda_k| (k=1,2,3)$ were 0.7, the variance of the indicators $\sigma^2_k$ was 1, the variance of the latent factor $\sigma^2_{\eta}$ was 1, and the intercepts were set at 0.

In the simulation, we varied the loadings in terms of their signs, creating four distinct conditions:\\
Condition 1: All loadings are positive.\\
Condition 2: All loadings are negative.\\
Condition 3: One of the three loadings is negative.\\
Condition 4: Two of the three loadings are negative.

The sample size was set at 200 for the simulation study. For each condition, 500 datasets were generated. We fitted the population model to each dataset in three popular CFA software packages, \textit{lavaan 0.6-16}, \textit{Mplus 8.10}, and \textit{OpenMx 2.21.8}.  These analyses used various fixing techniques. In the fixing factor variance approach, we used both the default starting values for factor loadings provided by the software packages and also manually set these starting values. Therefore, to thoroughly assess the impact of starting values on the direction of factor loadings, we used the default starting value, a positive starting value of 1, and a negative starting value of -1. For the fixed factor loading approach, the common default in most software is to fix the first-factor loading at 1. As a result, the four methods used in this simulation are: \\
Method 1: Fixing factor variance with default starting values of the factor loadings. \\
Method 2: Fixing factor variance with starting values of the factor loadings set to 1.\\
Method 3: Fixing factor variance with starting values of the factor loadings set to -1.\\
Method 4: Fixing the first loading at 1. 

\subsection{Results}
To evaluate the estimation problem with the four fixed methods, we calculated the directional consistency rate (DCR) to measure the possibility that the direction of the loading estimates is the same as the direction of the true values among 500 replicates. DCR can be expressed by $DCR = \frac{M}{N}*100\%$, where $M$ is the number of factor loading estimates with the same directional sign as the true values, and $N$ is the total number of factor loading estimates across all replicates. Table 2 presents the directional consistency rates for four fixing methods under various conditions in \textit{lavaan}. When using Method 4, where the first-factor loading is fixed at 1, DCR was not computed for this loading as it was fixed and cannot be freely estimated. Under various conditions,  the DCR for each loading was either 100\% or 0\%. A DCR of 100\% implies that across the replicates, the estimated factor loadings have the same directional sign (either all positive or all negative) as the true factor loadings, which is complete consistency. Conversely, a DCR of 0\% indicates that each estimated loading is directionally opposite to its true value, which is a complete lack of consistency. 

Method 1 could lead to loading estimates with inconsistent signs. For Methods 2 and 3, if the sign of the starting values was not consistent with true factor loadings, these loading estimates could have incorrect signs. For Method 4, the same directional inconsistency could arise when the first loading was fixed to 1, but its population value was negative. This implies that fixing a negative loading to 1 could mistakenly estimate positive loadings as negative or vice versa. Thus, the simulation findings underscore the potential drawbacks of these four fixing methods, particularly the risk of obtaining estimated loadings with unexpected signs.

We further validated our findings by running the simulation using  \textit{Mplus 8.10} and \textit{OpenMx 2.21.8}. Tables 3 and 4 display results largely consistent with those obtained using \textit{lavaan}. Although some difference is observed between Method 1 and 2, the overall results from \textit{Mplus} and \textit{OpenMx} further validate our findings that these four fixing methods may lead to loading estimates with inconsistent signs. Additionally, despite the unknown default starting values in these three software packages, Method 1 still can not guarantee the loading estimates in the expected directions.

\begin{table}[ht]
	\centering
	\caption{Directional consistency rates (\%) for simulation 1 with four fixing methods (\textit{lavaan})}
	\label{t2} 
	\begin{threeparttable}
	\begin{tabular}{rrrrrrrrrrrrr}
		\hline
          Fixing method & \multicolumn{3}{c}{Condition 1}  & \multicolumn{3}{c}{Condition 2} &\multicolumn{3}{c}{Condition 3} & \multicolumn{3}{c}{Condition 4} \\  
          \cline{1-13} 
          loading &  0.7 & 0.7 & 0.7 & -0.7 & -0.7 & -0.7 & -0.7 & 0.7 & 0.7  & -0.7  & -0.7  & 0.7  \\
             \hline
		 Method 1 &  100 & 100 & 100&  0 & 0 & 0&  0 & 0 & 0  &  0 & 0 & 0\\  
            Method 2 &  100 & 100 & 100&  0 & 0 & 0&  100 & 100 & 0  &  0 & 0 & 0\\
            Method 3 &  0 & 0 & 0&  100 & 100 & 100&  0 & 0 & 0  &  100 & 100 & 100\\
            Method 4 &   & 100 & 100&   & 0 & 0&   & 0 & 0  &   & 0 & 0\\
		\hline 
	\end{tabular}
	\begin{tablenotes}
        \footnotesize
      \end{tablenotes}
     \end{threeparttable}
\end{table}

\begin{table}[ht]
	\centering
	\caption{Directional consistency rates (\%) for simulation 1 with four fixing methods (\textit{Mplus})}
	\label{t3} 
	\begin{threeparttable}
	\begin{tabular}{rrrrrrrrrrrrr}
		\hline
          Fixed methods & \multicolumn{3}{c}{Condition 1}  & \multicolumn{3}{c}{Condition 2} &\multicolumn{3}{c}{Condition 3} & \multicolumn{3}{c}{Condition 4} \\   
          loading &  0.7 & 0.7 & 0.7 & -0.7 & -0.7 & -0.7 & -0.7 & 0.7 & 0.7  & -0.7  & -0.7  & 0.7  \\
             \hline 
            Method 1  &  100 & 100 & 100&  0 & 0 & 0&  0 & 0 & 0  &  0 & 0 & 0\\
            Method 2 &  100 & 100 & 100&  0 & 0 & 0&  0& 0& 0  &  0 & 0 & 0\\
            Method 3 &  0 & 0 & 0&  100 & 100 & 100&  0 & 0 & 0  &  100 & 100 & 100\\
            Method 4 &   & 100 & 100&   & 0 & 0&   & 0 & 0  &   & 0 & 100\\
		\hline 
	\end{tabular}
	\begin{tablenotes}
        \footnotesize
      \end{tablenotes}
     \end{threeparttable}
\end{table}

\begin{table}[ht]
	\centering
	\caption{Directional consistency rates (\%) for simulation 1 with four fixing methods (\textit{OpenMx})}
	\label{t4} 
	\begin{threeparttable}
	\begin{tabular}{rrrrrrrrrrrrr}
		\hline
          Fixed methods & \multicolumn{3}{c}{Condition 1}  & \multicolumn{3}{c}{Condition 2} &\multicolumn{3}{c}{Condition 3} & \multicolumn{3}{c}{Condition 4} \\   
          loading &  0.7 & 0.7 & 0.7 & -0.7 & -0.7 & -0.7 & -0.7 & 0.7 & 0.7  & -0.7  & -0.7  & 0.7  \\
             \hline 
            Method 1  &  100 & 100 & 100&  0 & 0 & 0&  100& 100& 100&  0& 0& 0\\
            Method 2 &  100 & 100 & 100&  0 & 0 & 0&  100 & 100 & 0  &  0 & 0 & 0\\
            Method 3 &  0 & 0 & 0&  100 & 100 & 100&  0 & 0 & 0  &  100 & 100 & 100\\
            Method 4 &   & 100 & 100&   & 0 & 0&   & 0 & 0  &   & 0 & 0 \\
		\hline 
	\end{tabular}
     \end{threeparttable}
\end{table}

\section{Simulation 2:  Solutions to the Estimation Problem}
\subsection{Simulation Design}
To address the issue of directional inconsistency in CFA factor loading estimates, we will propose two practical solutions and one potential solution that can yield estimates in a more interpretable and consistent direction. Furthermore, a Monte Carlo simulation was conducted to assess the effectiveness of these three solutions to the estimation problems for CFA. This simulation used the same population model as Simulation 1, with a sample size of 200 and 500 datasets generated for each condition.

\subsubsection{Solution 1: fixing a positive factor loading at 1}
Our previous simulation findings, particularly with Method 4, indicated that fixing a negative loading at 1 led to inconsistencies in the directions of other loadings. Conversely, fixing a positive loading at 1 aligned other loadings in the expected directions. Therefore, we used this solution in \textit{lavaan 0.6-16}, \textit{Mplus 8.10}, and \textit{OpenMx 2.21.8} for the population model under Conditions 1, 3, and 4, where not all loadings are negative. 

\subsubsection{Solution 2: setting bounds for factor loadings}
This solution is to fix the factor variance and set a lower bound for the inherently positive loadings and an upper bound for the inherently negative loadings. For instance, specifying 0 as the lower and upper bounds for positive and negative loadings, respectively. This solution is uniquely implementable in the OpenMx R package, which allows setting the same bound for all loading and a separate bound for each loading.  Consequently, we used this solution in \textit{OpenMx 2.21.8}, using 0 as the bound under Conditions 1 and 2, where each condition has the factor loadings with uniform signs. Specifically, we set the lower bound of 0 for all loadings under Condition 1 and an upper bound of 0 under Condition 2.

\subsubsection{Solution 3: adjusting starting values of factor loadings}
The potential solution is to fix the factor variance and set the starting values of the factor loadings in the same direction as the true loadings. As suggested by our simulation results for Methods 2 and 3, setting a start value consistent with the direction of true loadings could steer the estimates toward the desired direction. Therefore, we used this solution in \textit{lavaan 0.6-16}, \textit{Mplus 8.10}, and \textit{OpenMx 2.21.8} for each dataset under Conditions 1 to 4. Specifically, we set a large positive start value of 1 for a true positive loading or a small negative start value of -1 for a true negative loading. 

\subsection{Results}
The DCR was again computed to evaluate the effectiveness of three Solutions. Tables 5 to 7 present the DCR results for three solutions under various conditions. Solution 1, which fixes a positive factor loading at 1, successfully aligned other loading estimates in their correct direction. In Solution 2, fixing factor variance with a lower bound of 0 or an upper bound of 0 for all loadings effectively ensured that all loadings were consistently positive or negative. Solution 3, fixing the factor variance and setting the starting values to 1 or -1, proved effective in shifting the loadings toward the desired direction. From these findings, we concluded that Solutions 1 and 2 can solve the issue of directional inconsistency in factor loading, and Solution 3 can obtain the preferred direction of factor loadings.
\begin{table}[ht]
	\centering
	\caption{Directional consistency rates (\%) for Solution 1}
	\label{t5} 
	\begin{threeparttable}
	\begin{tabular}{rrrrrrrrrr}
		\hline
          Fixing method & \multicolumn{3}{c}{Condition 1}& \multicolumn{3}{c}{Condition 3} & \multicolumn{3}{c}{Condition 4} \\  
          \hline
          loading &  0.7 & 0.7 & 0.7 &  -0.7 & 0.7 & 0.7 & -0.7 & -0.7 & 0.7  \\
             \hline
		 Solution 1 (\textit{lavaan})  & 100& 100 &  & 100  & 100 &  &100  & 100 & \\  
         Solution 1 (\textit{Mplus}) & 100& 100 &  & 100  & 100 &  &100  & 100 & \\
         Solution 1 (\textit{OpenMx}) & 100& 100 &  & 100  & 100 &  &100  & 100 & \\
		\hline 
	\end{tabular}
	\begin{tablenotes}
        \footnotesize{Note: Under each condition, the last factor loading was fixed at 1.}
      \end{tablenotes}
     \end{threeparttable}
\end{table}

\begin{table}[ht]
	\centering
	\caption{Directional consistency rates (\%) for Solution 2}
	\label{t6} 
	\begin{threeparttable}
	\begin{tabular}{rrrrrrr}
		\hline
          Fixing method & \multicolumn{3}{c}{Condition 1} & \multicolumn{3}{c}{Condition 2} \\ 
          \hline
          loading &  0.7 & 0.7 & 0.7 & -0.7 & -0.7 & -0.7  \\
             \hline
             Solution 2 (\textit{OpenMX}) &  100 & 100 & 100 & 100 & 100 & 100\\    
		\hline 
	\end{tabular}
	\begin{tablenotes}
        \footnotesize
      \end{tablenotes}
     \end{threeparttable}
\end{table}

\begin{table}[ht]
	\centering
	\caption{Directional consistency rates (\%) for Solution 3}
	\label{t7} 
	\begin{threeparttable}
	\begin{tabular}{rrrrrrrrrrrrr}
		\hline
          Fixing method & \multicolumn{3}{c}{Condition 1}  & \multicolumn{3}{c}{Condition 2} &\multicolumn{3}{c}{Condition 3} & \multicolumn{3}{c}{Condition 4} \\   
          loading &  0.7 & 0.7 & 0.7 & -0.7 & -0.7 & -0.7 & -0.7 & 0.7 & 0.7  & -0.7  & -0.7  & 0.7  \\
             \hline 
            Solution 3 (\textit{lavaan}) &  100 & 100 & 100&  100 & 100 & 100&  100 & 100 & 100  &  100 & 100 & 100\\
             Solution 3 (\textit{Mplus}) &  100 & 100 & 100&  100 & 100 & 100&  100 & 100 & 100  &  100 & 100 & 100\\
              Solution 3 (\textit{OpenMX}) &  100 & 100 & 100&  100 & 100 & 100&  100 & 100 & 100  &  100 & 100 & 100\\
		\hline 
	\end{tabular}
     \end{threeparttable}
\end{table}

\section{Reanalysis of the Empirical Example with the Solutions}
In practice, we can use the three solutions to handle the issue of directional inconsistency in factor loading estimates. However, not all proposed solutions can be conducted in popular CFA software packages. Specifically, \textit{lavaan} and \textit{OpenMx} support the first solution for both continuous and categorical variables, while \textit{Mplus} only supports continuous variables. The second solution can only be applied in \textit{OpenMx}. The third solution can be used in these three software. However, \textit{OpenMx} can not handle more than 20 categorical observed variables. Detailed implementation guidelines and codes are available in the Appendix 1.

Since there are more than 20 categorical observed variables in the empirical example, the second solution can not be applied in this example. Only Solutions 1 and 3 are used through \textit{lavaan} and \textit{Mplus} (see the codes in our href{https://github.com/DandanTang0/sign-of-factor-loading/tree/main/example}{GitHub}
). For Solution 1, the first-factor loading is fixed at 1 because the first item, Mood (Questions related to feelings of sadness or happiness) and the latent factor of Children's Depression have a positive relationship given substantive psychological theory. For Solution 3,  a positive start value of 1 is applied because all the items have a positive relationship with the latent factor of Children's Depression. The re-analysis results are presented in Table 8. All loadings at T1 in the data without missingness are positive. This implies that Solutions 1 and 3 work for this example and the previous non-replicated results were due to how CFA was implemented.  

\begin{table}[ht]
	\centering
	\caption{Factor loadings for a one-factor model of the CDI at Time 1 in the data without missingness}
	\label{t8} 
	\begin{threeparttable}
	\begin{tabular}{rrrrr}
		\hline
        Item & \multicolumn{2}{c}{\textit{lavaan}}  & \multicolumn{2}{c}{\textit{Mplus}} \\ 
        \cline{2-5}
        & Solution 1  & Solution 3  & Solution 1 & Solution 3\\ 
             \hline
		CDI1 &  1.000  & 0.773 & 1.000 & 0.775\\ 
		CDI2 &   0.827  & 0.639 & 0.820& 0.635\\ 
		CDI3 &  0.795 & 0.614 & 0.793 & 0.615\\ 
            CDI4 &  0.679  & 0.525 & 0.675& 0.523\\ 
            CDI5 &  0.733  & 0.567 & 0.732& 0.567\\ 
            CDI6 &  0.600  & 0.464 & 0.614&0.476\\ 
            CDI7 &  0.967  & 0.747 & 0.898&0.696\\ 
            CDI8 &  0.824  & 0.637 & 0.822&0.637\\ 
            CDI9 &  0.892 & 0.690 & 0.890 &0.689\\ 
            CDI10 &  1.064  & 0.823 & 1.062&0.823\\ 
            CDI11 &  0.990 & 0.765 & 1.002& 0.777\\ 
            CDI12 &  0.747  & 0.577 & 0.717& 0.555\\ 
            CDI13 &  0.689  & 0.532 & 0.685& 0.531\\ 
            CDI14 &  0.701  & 0.541 & 0.694 & 0.538\\ 
            CDI15 &  0.592  & 0.457 & 0.592& 0.459\\ 
            CDI16 &  0.833  & 0.644 & 0.832 & 0.645\\ 
            CDI17 &  0.898  & 0.694 & 0.894 & 0.693\\ 
            CDI18 &  0.522  & 0.404 & 0.531& 0.412\\ 
            CDI19 &  0.516  & 0.399 & 0.500& 0.388\\ 
            CDI20 &  1.053  & 0.814 & 1.050& 0.813 \\  
            \vdots &  \vdots  & \vdots & \vdots & \vdots\\ 
            CDI27 &  0.895  & 0.691 & 0.894 & 0.693\\ 
		\hline 
	\end{tabular}
	\begin{tablenotes}
        \footnotesize{Note: The table only lists part of the results because of limited space. For complete results, please see our \href{https://github.com/DandanTang0/sign-of-factor-loading/tree/main/table}{GitHub}.
        }
      \end{tablenotes}
     \end{threeparttable}
\end{table}

\section{Prevalence of the Issue and Practical Suggestions}
The issue of estimating factor loading direction in CFA models has often been overlooked in practice. Yet, its significance cannot be overstated, especially in comparative and longitudinal studies. Comparative research requires close attention to factor loading signs, particularly in measurement invariance tests. As the practical example shows, ignoring the sign of factor loadings makes it difficult to determine whether the loadings are invariant across groups or over time. Furthermore, in longitudinal research, which tracks individual changes over time (Fitzmaurice et al., 2012), inconsistent loading signs may distort true within-individual changes. Additionally, the issue extends to cross-sectional studies where mismatched loading signs can make it difficult to validate the structure of a scale (Lin et al., 2019).

In general, any study using CFA should consider the problem of factor loading signs in case the results of this study can not be replicated due to loading signs. Furthermore, this estimation problem is not confined to CFA models but is also prevalent in other structural equation models, as they employ similar estimation techniques. It is imperative to address the issue in all SEM models to avoid a potential replication dilemma.

To address the problem of loading signs, we proposed three solutions to reflect the true relationship between observed variables and latent factors. Even if this article only discussed it in CFA models, the solutions can be applied in all SEM models. The first solution, fixing a positive loading at 1,  can ensure that other loadings in the model are in the expected direction. Using substantive theory, researchers can identify an observed variable with a positive correlation with a latent factor and fix the loading on this variable to 1. However, researchers must identify and fix one positive loading for each latent factor. The second solution, fixing the factor variance and setting a  bound for loadings, can ensure that all loadings are in the correct direction. However, to set the right bound for each loading, researchers must determine all the relationships between observed variables and latent factors. The third solution,  fixing the factor variance and setting the starting values of loadings at larger values with the same direction as the loadings,  leads the loadings to the desired direction but does not guarantee it. Like the second solution, it requires a comprehensive understanding of the relationships between observed variables and latent factors.
\section{Author contributions}
Dandan Tang: Formal Analysis, Investigation, Methodology, Visualization, Writing-original draft;
Steven M. Boker: Methodology, Supervision, Writing-review \& editing; Xin Tong: Supervision, Writing-review \& editing.
\section{Acknowledgements}
The authors gratefully thank Dr. Caspar van Lissa for the valuable guidance on the empirical example.

\section{References}
\begin{description}
\item Anderson, S. F., \& Maxwell, S. E. (2017). Addressing the "replication crisis": Using original studies to design replication studies with appropriate statistical power. \emph{Multivariate Behavioral Research, 52}(3), 305-324.
\item Ali, Z., \& Bhaskar, S. B. (2016). Basic statistical tools in research and data analysis.\emph{Indian journal of anaesthesia, 60}(9), 662.
\item Asparouhov, T., \& Muthen, B. (2007, August). Computationally efficient estimation of multilevel high-dimensional latent variable models. \emph{In proceedings of the 2007 JSM meeting in Salt Lake City, Utah, Section on Statistics in Epidemiology} (pp. 2531-2535).
\item Arora, J. S., Elwakeil, O. A., Chahande, A. I., \& Hsieh, C. C. (1995). Global optimization methods for engineering applications: a review. \emph{Structural optimization, 9}, 137-159.
\item Bargh, J. A., Chen, M., \& Burrows, L. (1996). Automaticity of social behavior: Direct effects of trait construct and stereotype activation on action. \textit{Journal of personality and social psychology}, \textit{71}(2), 230.
\item Boker, S., Neale, M., Maes, H., Wilde, M., Spiegel, M., Brick, T., ... \& Fox, J. (2011). OpenMx: an open source extended structural equation modeling framework. \emph{Psychometrika, 76}, 306-317.
\item Camerer, C. F., Dreber, A., Holzmeister, F., Ho, T. H., Huber, J., Johannesson, M., ... \& Wu, H. (2018). Evaluating the replicability of social science experiments in Nature and Science between 2010 and 2015. Nature human behaviour, 2(9), 637-644.
\item Cockburn, A., Dragicevic, P., Besançon, L., \& Gutwin, C. (2020). Threats of a replication crisis in empirical computer science. \emph{Communications of the ACM, 63}(8), 70-79.
\item Crane, H. (2018). The impact of P-hacking on "redefine statistical significance". \emph{Basic and Applied Social Psychology, 40}(4), 219-235.
\item Fitzmaurice, G. M., Laird, N. M., \& Ware, J. H. (2012). \emph{Applied longitudinal analysis}, 2nd ed. John Wiley \& Sons.
\item Franco, A., Malhotra, N., \& Simonovits, G. (2014). Publication bias in the social sciences: Unlocking the file drawer. \textit{Science}, \textit{345}(6203), 1502-1505.
\item Harris, C. R., Coburn, N., Rohrer, D., \& Pashler, H. (2013). Two failures to replicate high-performance-goal priming effects. \textit{PloS one}, \textit{8}(8), e72467.
\item Klein, R. A., Vianello, M., Hasselman, F., Adams, B. G., Adams Jr, R. B., Alper, S., ... \& Sowden, W. (2018). Many Labs 2: Investigating variation in replicability across samples and settings. \textit{Advances in Methods and Practices in Psychological Science}, \textit{1}(4), 443-490.
\item Li, C. H. (2016). Confirmatory factor analysis with ordinal data: Comparing robust maximum likelihood and diagonally weighted least squares. \emph{Behavior research methods, 48}, 936-949.
\item Lin, X. F., Tang, D., Lin, X., Liang, Z. M., \& Tsai, C. C. (2019). An exploration of primary school students’ perceived learning practices and associated self-efficacies regarding mobile-assisted seamless science learning. \textit{International Journal of Science Education}, \textit{41}(18), 2675-2695.
\item Lin, X. F., Tang, D., Shen, W., Liang, Z. M., Tang, Y., \& Tsai, C. C. (2020). Exploring the relationship between perceived technology-assisted teacher support and technology-embedded scientific inquiry: the mediation effect of hardiness. \textit{International Journal of Science Education}, \textit{42}(8), 1225-1252.
\item Liu, H., Depaoli, S., \& Marvin, L. (2022a). Understanding the deviance information criterion for sem: Cautions in prior specification. \textit{Structural Equation Modeling: A Multidisciplinary Journal}, \textit{29}(2), 278-294.
\item Liu, H., Qu, W., Zhang, Z., \& Wu, H. (2022b). A New Bayesian Structural Equation Modeling Approach with Priors on the Covariance Matrix Parameter. \textit{Journal of Behavioral Data Science}, \textit{2}(2), 23-46.  
\item Kochenderfer, M. J., \& Wheeler, T. A. (2019). \emph{Algorithms for optimization}. Mit Press.
\item Muthén, L. K., \& Muthén, B. (2017). \emph{Mplus user's guide: Statistical analysis with latent variables}
\item Muthukrishna, M., \& Henrich, J. (2019). A problem in theory. \textit{Nature Human Behaviour}, \textit{3}(3), 221-229.
\item Nosek, B. A., Hardwicke, T. E., Moshontz, H., Allard, A., Corker, K. S., Dreber, A., ... \& Vazire, S. (2022). Replicability, robustness, and reproducibility in psychological science. \textit{Annual review of psychology}, \textit{73}, 719-748.
\item Oberauer, K., \& Lewandowsky, S. (2019). Addressing the theory crisis in psychology. \emph{Psychonomic bulletin \& review, 26}, 1596-1618.
\item Open Science Collaboration. (2015). Estimating the reproducibility of psychological science. \emph{Science, 349}(6251), aac4716.
\item Pashler, H., Coburn, N., \& Harris, C. R. (2012). Priming of social distance? Failure to replicate effects on social and food judgments.
\item Rosseel, Y. (2012). \emph{lavaan: a brief user's guide}. 
\item Smaldino, P. E., \& McElreath, R. (2016). The natural selection of bad science. \textit{Royal Society open science}, \textit{3}(9), 160384.
\item Stumper, A., Olino, T. M., Abramson, L. Y., \& Alloy, L. B. (2019). A factor analysis and test of longitudinal measurement invariance of the Children's Depression Inventory (CDI) across adolescence. \textit{Journal of psychopathology and behavioral assessment}, \textit{41}, 692-698.
\item Tang, D., \& Tong, X. (2023). A Comparison of Full Information Maximum Likelihood and Machine Learning Missing Data Analytical Methods in Growth Curve Modeling. \textit{arXiv preprint arXiv:2312.17363}.
\item Van Lissa, C. J., Brandmaier, A. M., Brinkman, L., Lamprecht, A. L., Peikert, A., Struiksma, M. E., \& Vreede, B. M. (2021). WORCS: A workflow for open reproducible code in science. \emph{Data Science, 4}(1), 29-49.
\item Vandenberg, R. J., \& Lance, C. E. (2000). A Review and Synthesis of the Measurement Invariance Literature: Suggestions, Practices, and Recommendations for Organizational Research. \emph{Organizational Research Methods, 3(1)}, 4–70. https://doi.org/10.1177/109442810031002
\item Wagner III, J. A. (2022). The influence of unpublished studies on results of recent meta-analyses: Publication bias, the file drawer problem, and implications for the replication crisis. \emph{International Journal of Social Research Methodology, 25}(5), 639-644.
\item WEN, Z., TANG, D., \& GU, H. (2019). A general simulation comparison of the predictive validity between bifactor and high-order factor models. \textit{Acta Psychologica Sinica}, \textit{51}(3), 383.
\item Youyou, W., Yang, Y., \& Uzzi, B. (2023). A discipline-wide investigation of the replicability of Psychology papers over the past two decades. \emph{Proceedings of the National Academy of Sciences, 120}(6), e2208863120.
\item Zhong-Lin, Y. U. N., Ban-Ban, H. U. A. N. G., \& Dan-Dan, T. A. N. G. (2018). Preliminary work for modeling questionnaire data. \textit{Journal of Psychological Science}, \textit{41}(1), 204.
\item Zhonglin, D. T. W. (2020). Statistical approaches for testing common method bias: Problems and suggestions. \textit{Journal of Psychological Science}, (1), 215.
\end{description}

\subsection{Appendix 1}
\subsubsection{Continuous variables}
Suppose a one-factor model with three continuous variables. This is the R code for \textbf{Solution 1}, a positive loading at 1 using \textit{lavaan} package. If the true value of the first loading is positive,  the first loading can be fixed at 1 by "auto.fix.first = TRUE"
\begin{lstlisting}[language=R]
# load package
library(lavaan)
# read data
data <- 
    read.csv('data.csv',header = T)
# CFA model for T1
CFA <- 'F=~x1+x2+x3'
# The true value of the first loading is 
# positive, so the first loading is 
# fixed at 1 by "auto.fix.first = TRUE" 
result <- cfa(CFA, data = data,
auto.fix.first = TRUE, ordered = FALSE)
# print result
summary(result)
\end{lstlisting}
If the first loading is not positive, we can move a positive loading to the first place. for example, if the loading on item x2 is positive, we can set the CFA model as follows:
\begin{lstlisting}[language=R]
CFA_reloaction <- 'F=~x2+x1+x3'
\end{lstlisting}
This is the R code for \textbf{Solution 3} using \textit{lavaan} package. The factor variance can be fixed by "std.lv = TRUE", and the starting values of the loadings can be set at 1 or -1 by "start(1)*" or "start(-1)*".
\begin{lstlisting}[language=R]
# set start values of the 
# loadings at 1 by "start(1)*"
CFA_start <- 'F=~start(1)*x1+start(1)*x2
+start(1)*x3'
# fix the factor variance by "std.lv = TRUE" 
result <- cfa(CFA_start, data = data,
std.lv = TRUE, ordered = FALSE)
# print results
summary(result)
\end{lstlisting}
This is \textit{Mplus} code for \textbf{Solution 1}, fixing a positive loading at 1. If the true value of the first loading is positive,  the first loading is fixed at 1 by "@1".
\begin{lstlisting}[language=R]
 ! read data
 DATA: FILE = data.dat; 
! name observed variables
  VARIABLE: NAMES =  x1-x3; 
  Model:
! fix the first loading at 1 by "@1"
       F by x1@1 x2 x3; 
\end{lstlisting}
If the first loading is not positive, we can move a positive loading to the first place. for example, if the loading on item x2 is positive, we can set its loadings at 1 by "@1".
\begin{lstlisting}[language=R]
DATA: FILE = data.dat;
  VARIABLE: NAMES =  x1-x3;
  Model:
! fix the second loading at 1 by "@1"
        F by x1 x2@1 x3; 
\end{lstlisting}
This is \textit{Mplus} code for \textbf{Solution 3}, fixing the factor variance by "@1" and setting the starting values of the loadings to 1 or -1 by "*1" or "*-1".
\begin{lstlisting}[language=R]
DATA: FILE = data.dat;
  VARIABLE: NAMES =  x1-x3;
  Model:
! set the starting values of 
! the loadings to 1 by "*1"
        F by x1-x3*1; 
! fix the factor variance by "@1"
           F@1; 
\end{lstlisting}
This is the R code for \textbf{Solution 1}, a positive loading at 1 using \textit{OpenMx} package.  If the true value of the first loading is positive,  the first loading is fixed at 1 by "mxPath(from=c("F"), to=c("x1"), arrows=1, free=FALSE, values=1)".
\begin{lstlisting}[language=R]
# load package
library(OpenMx)
# read data
data <- read.csv("data.csv")
# name variables and parameters
indicators <- names(data)
latents <- c("F")
loadingLabels <- 
   paste("b_",indicators, sep="")
uniqueLabels <- 
   paste("U_",indicators, sep="")
meanLabels <- 
   paste("M_", indicators, sep="")
factorVarLabels <- 
   paste("Var_", latents, sep="")
# build model
oneFactorRaw1 <- mxModel(
"Single factor Model with Fixed Loading",
type= "RAM",
manifestVars=indicators,
latentVars=latents,
mxPath(from=latents, to=indicators, 
arrows=1, connect= "unique.bivariate",
free=TRUE, values=.2, 
labels=loadingLabels),
mxPath(from=c("F"), to=c( "x1"), 
arrows=1, free=FALSE, values=1),
mxPath(from=indicators, 
arrows=2, 
free=TRUE, values=.8,
labels=uniqueLabels),
mxPath(from=latents,
arrows=2,
free=TRUE, values=.8, 
labels=factorVarLabels),
mxPath(from="one", to=indicators, 
arrows=1, free=TRUE, values=.1, 
labels=meanLabels),
mxData(observed=data, type="raw")
)
oneFactorRaw1Out <- mxRun(oneFactorRaw1)
# print results
summary(oneFactorRaw1Out)
\end{lstlisting}
If the first loading is not positive, we can move a positive loading to the first place. for example, if the loading on item x2 is positive, we can set its loadings at 1 by "mxPath(from=c("F"), to=c("x2"), arrows=1, free=FALSE, values=1)".
\begin{lstlisting}[language=R]
# build model
oneFactorRaw1 <- mxModel(
"Single factor Model with Fixed Loading",
type= "RAM",
manifestVars=indicators,
latentVars=latents,
mxPath(from=latents, to=indicators, 
arrows=1, connect= "unique.bivariate",
free=TRUE, values=.2, 
labels=loadingLabels),
mxPath(from=c("F"), to=c( "x2"), 
arrows=1, free=FALSE, values=1),
mxPath(from=indicators, 
arrows=2, 
free=TRUE, values=.8,
labels=uniqueLabels),
mxPath(from=latents,
arrows=2,
free=TRUE, values=.8, 
labels=factorVarLabels),
mxPath(from="one", to=indicators, 
arrows=1, free=TRUE, values=.1, 
labels=meanLabels),
mxData(observed=data, type="raw")
)
\end{lstlisting}

This is the R code for \textbf{Solution 3} using \textit{OpenMx} package. The factor variance is fixed to 1 by setting "free=FALSE and values=1" and the starting values of the loadings are set at 1 by "free=TRUE and  values=1".
\begin{lstlisting}[language=R]
# build model
oneFactorRaw1 <- mxModel(
"Single Factor model with Fixed Variance", 
type= "RAM",
manifestVars=indicators,
latentVars=latents,
mxPath(from=latents, to=indicators, 
arrows=1, connect= "unique.bivariate", 
# set the starting values of 
# the loadings at 1.
free=TRUE, values=1, 
labels=loadingLabels),
mxPath(from=indicators, 
arrows=2, 
free=TRUE, values=.8, 
labels=uniqueLabels),
mxPath(from=latents,
arrows=2, 
# fixing factor variance to 1.
free=FALSE, values=1, 
labels=factorVarLabels),
mxPath(from="one", to=indicators, 
arrows=1, free=TRUE, values=.1, 
labels=meanLabels),
mxData(observed=data, type="raw")
    )
\end{lstlisting}
the R code for \textbf{Solution 2}, a positive loading at 1 using \textit{OpenMx} package. We can set the lower bound of 0 by "lbound=0", or the upper bound of 0 by "ubound=0"
\begin{lstlisting}[language=R]
oneFactorRaw1 <- mxModel(
"Single Factor model with Fixed Variance", 
type= "RAM",
manifestVars=indicators,
latentVars=latents,
mxPath(from=latents, to=indicators, 
arrows=1, connect= "unique.bivariate", 
# setting the lower 
# bound of the loadings >0.
free=TRUE, lbound=0, 
labels=loadingLabels),
mxPath(from=indicators, 
arrows=2, 
free=TRUE, values=.8, 
labels=uniqueLabels),
mxPath(from=latents,
arrows=2, 
# fixing factor variance to 1.
free=FALSE, values=1, 
labels=factorVarLabels),
mxPath(from="one", to=indicators, 
arrows=1, free=TRUE, values=.1, 
labels=meanLabels),
mxData(observed=data, type="raw")
    )
\end{lstlisting}
\subsubsection{Categorical variables}
Suppose a one-factor model with three categorical variables. If using \textit{lavaan}, "ordered = FALSE" should be changed into "ordered = TRUE" and the rest part should be kept the same.  

This is \textit{Mplus} code for \textbf{Solution 1}, fixing a positive loading at 1. If the true value of the first loading is positive,  the first loading is fixed at 1 by "@1".
\begin{lstlisting}[language=R]
DATA: FILE = data.dat;
  VARIABLE: NAMES =  x1-x3;
   CATEGORICAL ARE x1-x3;
  ! The estimation method is WLSMV
  ANALYSIS: ESTIMATOR = WLSMV; 
  Model:
 !! fix the first loading at 1 by "@1"
        F by x1@1 x2-x3*1;
OUTPUT: TECH1 TECH8;
\end{lstlisting}
This is \textit{Mplus} code for \textbf{Solution 3}, fixing the factor variance by "@1" and setting the starting values of the loadings to 1 or -1 by "*1" or "*-1".
\begin{lstlisting}[language=R]
DATA: FILE = data.dat;
  VARIABLE: NAMES =  x1-x3;
   CATEGORICAL ARE x1-x3;
  ! The estimation method is WLSMV
  ANALYSIS: ESTIMATOR = WLSMV; 
  Model:
 !set the starting values of 
 ! the loadings to 1 by "*1"
        F by x1-x3*1; 
 ! fix the factor variance at 1 by "@1"
        F @1;
OUTPUT: TECH1 TECH8;
\end{lstlisting}

\end{document}